\documentclass[a4paper,twoside,twocolumn,english,aps,superscriptaddress,showpacs]{revtex4}
\usepackage{times}
\usepackage[T1]{fontenc}
\usepackage[latin1]{inputenc}
\usepackage{subfigure}
\usepackage{amsmath}
\usepackage{color}
\usepackage{graphicx}
\usepackage{amssymb}

\makeatletter

\newcommand{\noun}[1]{\textsc{#1}}
\newcommand{\lyxdot}{.}

\usepackage{babel}
\makeatother
\begin{document}

\title{Tunneling dynamics of few bosons in a double well}

\author{Sascha Zöllner}

\email{sascha.zoellner@pci.uni-heidelberg.de}

\affiliation{Theoretische Chemie, Universit\"{a}t Heidelberg, Im Neuenheimer
Feld 229, 69120 Heidelberg, Germany}

\author{Hans-Dieter Meyer}

\affiliation{Theoretische Chemie, Universit\"{a}t Heidelberg, Im Neuenheimer
Feld 229, 69120 Heidelberg, Germany}

\author{Peter Schmelcher}

\email{peter.schmelcher@pci.uni-heidelberg.de}

\affiliation{Theoretische Chemie, Universit\"{a}t Heidelberg, Im Neuenheimer
Feld 229, 69120 Heidelberg, Germany}

\affiliation{Physikalisches Institut, Universit\"{a}t Heidelberg, Philosophenweg
12, 69120 Heidelberg, Germany}

\begin{abstract}
We study few-boson tunneling in a one-dimensional double well. As
we pass from weak interactions to the fermionization limit, the Rabi
oscillations first give way to highly delayed pair tunneling (for
medium coupling), whereas for very strong correlations multi-band
Rabi oscillations emerge. All this is explained on the basis of the
exact few-body spectrum and without recourse to the conventional two-mode
approximation. Two-body correlations are found essential to the understanding
of the different tunnel mechanisms. The investigation is complemented
by discussing the effect of skewing the double well, which offers
the possibility to access specific tunnel resonances.
\end{abstract}

\date{January 4, 2008}


\maketitle

\section{Introduction}

Using ultracold atoms, it has become possible to study hallmark quantum
effects---such as tunneling---at an unprecedented level of precision
and control \cite{pitaevskii,dalfovo99,pethick,leggett01}. One prime
example is tunneling of matter waves, where Bose-Einstein condensates
have facilitated the observation of Josephson oscillations \cite{albiez05,milburn97,smerzi95}
and the complementary nonlinear self-trapping \cite{albiez05,anker05,javanainen86}.
In the case of Josephson oscillations, the atoms---initially prepared
mostly in one well---simply tunnel back and forth between two potential
wells in analogy to a current in a Josephson junction. However, above
a critical interaction strength, the atoms essentially remain trapped
in that well for the experimental lifetime even though they repel
each other. 

While the above effects have been observed for macroscopic coherent
matter waves, many tools such as optical lattices have promoted a
trend to study smaller systems with few atoms only. Permitting a high
degree of control, they offer the chance to study finite-size effects
and this way allow for a deeper understanding of the microscopic mechanisms
in ultracold atoms. As an example, the recently evidenced stability
of repulsively interacting atom pairs as they move in a lattice \cite{winkler06},
as well as the direct observation of their first- and second-order
tunneling dynamics \cite{foelling07}, should be seen as few-body
counterparts of the above self-trapping transition. This motivates
a thorough theoretical investigation of the few-boson tunneling mechanisms.

However, while those effects are confined to the regime of relatively
weak interactions, interatomic forces can be adjusted experimentally
over a wide range, e.g., by exploiting Feshbach resonances \cite{koehler06}.
In particular, it is well known that in one dimension (1D) one can
tune the effective interaction strength at will via a confinement-induced
resonance \cite{Olshanii1998a}, which makes it possible to explore
the limit of strong correlations. If the bosons repel each other infinitely
strongly, they can be mapped to noninteracting fermions \cite{girardeau60},
in that the exclusion principle serves to mimic the \emph{hard-core}
interaction. While the bosons share local aspects with their fermionic
counterparts, nonlocal properties such as their momentum distribution
are very different. Sparked also by the experimental demonstration
\cite{paredes04,kinoshita04,kinoshita06}, this \emph{fermionization}
has attracted broad interest (see \cite{girardeau00,das02,Busch03,minguzzi05,alon05,deuretzbacher06,zoellner06a,zoellner06b,zoellner07a}
and Refs. therein).

In this light, the question naturally arises whether the notion of
tunneling can be pushed to the strongly interacting fermionization
limit. Indeed, a recent study has shown that a fermionized atom pair
tunnels coherently almost like a single atom \cite{zoellner07c}.
In this paper, we give a systematic account of how few-boson tunneling
evolves in the crossover from weak to strong correlations. Moreover,
we extend that study to two-atom tunneling resonances occuring in
asymmetric wells.

Our paper is organized as follows. Section~\ref{sec:theory} introduces
the model and briefly reviews the concept of fermionization. In Sec.~\ref{sec:method},
we give a concise presentation of the computational method. The subsequent
section is devoted to the results on tunneling in a symmetric double
well for two atoms (Secs.~\ref{sub:Two-atoms}--\ref{sub:correlations})
and more atoms (Sec.~\ref{sub:Many-body}). Finally, we illuminate
the effect of tilting the double well in Sec.~\ref{sec:Asymmetric}.

\section{Theoretical background \label{sec:theory}}

\subsection{Model}

The subject of this article is the double-well dynamics of few atoms
($N=2-4$), which shall be described by the many-body Hamiltonian
(see \cite{zoellner06a} for details) \[
H=\sum_{i=1}^{N}\left[\frac{1}{2}p_{i}^{2}+U(x_{i})\right]+g\sum_{i<j}\delta_{\sigma}(x_{i}-x_{j}).\]
Here the double-well trap $U(x)=\frac{1}{2}x^{2}+h\delta_{w}(x)$
is modeled as a superposition of a harmonic oscillator and a central
barrier shaped as a Gaussian $\delta_{w}(x)=e^{-x^{2}/2w^{2}}/\sqrt{2\pi}w$
(of width $w=0.5$, where harmonic-oscillator units are employed throughout.)
The effective interaction in 1D can be represented as a contact potential
\cite{Olshanii1998a}, but is mollified here with a Gaussian $\delta_{\sigma=0.05}$
 so as to alleviate the well-known numerical difficulties caused
by the $\delta$ function. We focus on repulsive forces, i.e., $g\in[0,\infty)$.

To prepare the initial state $\Psi(0)$ with a population imbalance---in
our case, such that almost all atoms reside in the \emph{right} well
only---we make that side energetically favorable by adding a linear
external potential $-d\cdot x$ (with sufficiently large $d\sim0.1-1$,
depending on $N$ and $g$) and let the system relax to its ground
state $\Psi_{0}^{(d>0)}$. To study the time evolution in the \emph{symmetric}
double well (Sec.~\ref{sec:Symmetric}), the asymmetry $d$ will
be ramped down to $d\to0$ nonadiabatically (we typically choose a
ramp time $\tau\sim1$).  By extension, it is possible to take any
final asymmetry $\lim_{t\to\infty}d(t)\neq0$, which allows us to
look at the case where one well is energetically offset (Sec.~\ref{sec:Asymmetric}).

\subsection{Fermionization}

A peculiarity of 1D systems is that bosons with infinitely strong
repulsive point interactions, $g\to\infty$, become impenetrable.
Mathematically, this means that its configuration space becomes disconnected
into regions $\{ x_{i}\neq x_{j}\:\forall i<j\}$, a feature which
allows the system to be solved exactly via the \emph{Bose-Fermi map}
\cite{girardeau60} that establishes an isomorphy between the exact
\emph{bosonic} wave function $\Psi_{g\to\infty}^{+}$ and that of
a (spin-polarized) non-interacting \emph{fermionic} solution $\Psi_{0}^{-}$,
\begin{equation}
\Psi_{\infty}^{+}=A\Psi_{0}^{-},\label{eq:BF}\end{equation}
where $A=\prod_{i<j}\mathrm{sgn}(x_{i}-x_{j})$. The mapping rests
on general grounds and is valid for both stationary and explicitly
time-dependent states. Since $A^{2}=1$, their (diagonal) densities
as well as their energy $E$ will coincide with those of the corresponding
free fermionic states. That makes it tempting to think of the exclusion
principle as mimicking the interaction ($g\to\infty$), which is why
this limit is commonly referred to as \emph{fermionization}.

\section{Computational method\label{sec:method}}

Our goal is to investigate the few-atom quantum dynamics in the crossover
to the highly correlated fermionization limit $g\to\infty$ in an
\emph{exact} fashion. This is numerically challenging, and most studies
on the double-well dynamics so far have relied on two-mode models
\cite{milburn97,tonel05,salgueiro06,creffield07,dounasfrazer07a}
valid for sufficiently weak coupling. Here we adopt the Multi-Configuration
Time-Dependent Hartree \noun{(mctdh)} method \cite{mey90:73,bec00:1,mctdh:package}.
Its principal idea is to solve the time-dependent Schrödinger equation\[
\begin{array}{c}
i\dot{\Psi}(t)=H\Psi(t)\end{array}\]
 as an initial-value problem by expanding the solution in terms of
direct (or Hartree) products $\Phi_{J}\equiv\varphi_{j_{1}}\otimes\cdots\otimes\varphi_{j_{N}}$:\begin{equation}
\Psi(t)=\sum_{J}A_{J}(t)\Phi_{J}(t).\label{eq:mctdh-ansatz}\end{equation}
The (unknown) \emph{single-particle functions} $\varphi_{j}$ ($j=1,\dots,n$)
are in turn represented in a fixed \emph{primitive} basis implemented
on a grid. 

Note that in the above expansion not only the coefficients $A_{J}$
but also the single particle functions $\varphi_{j}$ are time dependent.
Using the Dirac-Frenkel variational principle, one can derive equations
of motion for both $A_{J},\varphi_{j}$ \cite{bec00:1}. Integrating
this differential-equation system allows us to obtain the time evolution
of the system via (\ref{eq:mctdh-ansatz}). This has the advantage
that the basis $\{\Phi_{J}(t)\}$ is variationally optimal at each
time $t$. Thus it can be kept relatively small, rendering the procedure
very efficient. We stress that $\Psi$ obeys bosonic permutation symmetry
even though the direct-product basis does not; this is ensured by
correct symmetrization of the expansion coefficients.

Although designed for time-dependent simulations, it is also possible
to apply this approach to stationary states. This is done via the
so-called \emph{relaxation method} \cite{kos86:223}. The key idea
is to propagate some wave function $\Psi(0)$ by the non-unitary $e^{-H\tau}$
(\emph{propagation in imaginary time}.) As $\tau\to\infty$, this
exponentially damps out any contribution but that stemming from the
true ground state like $e^{-(E_{m}-E_{0})\tau}$. In practice, one
relies on a more sophisticated scheme termed \emph{improved relaxation}
\cite{mey03:251,meyer06}, which is much more robust especially for
excitations. Here $\langle\Psi|H|\Psi\rangle$ is minimized with respect
to both the coefficients $A_{J}$ and the orbitals $\varphi_{j}$.
The equations of motion thus obtained are then solved iteratively
by first solving for $A_{J}$ with \emph{fixed} orbitals and then
{}`optimizing' $\varphi_{j}$ by propagating them in imaginary time
over a short period. That cycle will then be repeated.

\section{Symmetric double well \label{sec:Symmetric}}

Let us first focus on the tunnel dynamics in a symmetric well ($d=0$).
Our primary focus is on how the tunneling changes as we pass from
single-particle---i.e., uncorrelated---tunneling ($g=0$) to tunneling
in the presence of correlations and finally to the fermionization
limit ($g\to\infty$). It is natural to first look at the conceptually
clearest situation where $N=2$ atoms initially reside in the right-hand
well (Sec.~\ref{sub:Two-atoms}), with an eye toward the link between
tunneling times and the few-body spectrum (Sec.~\ref{sub:Spectrum})
as well as the role of two-body correlations (Sec.~\ref{sub:correlations}).
With this insight, we tackle the more complicated dynamics of $N=3,4,\dots$
atoms in Sec.~\ref{sub:Many-body}.

\subsection{From uncorrelated to pair tunneling \label{sub:Two-atoms}}

\begin{figure}
\includegraphics[width=0.85\columnwidth,keepaspectratio]{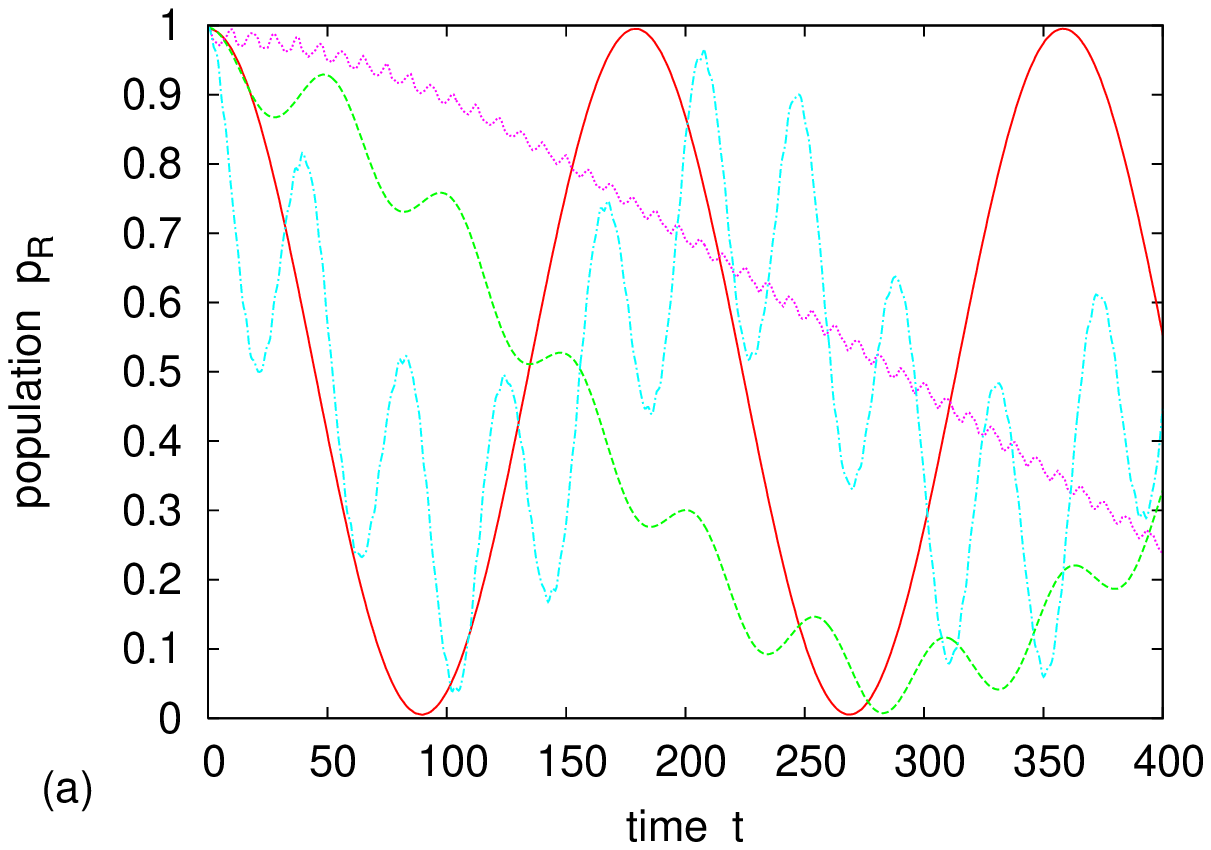}

\includegraphics[width=0.85\columnwidth,keepaspectratio]{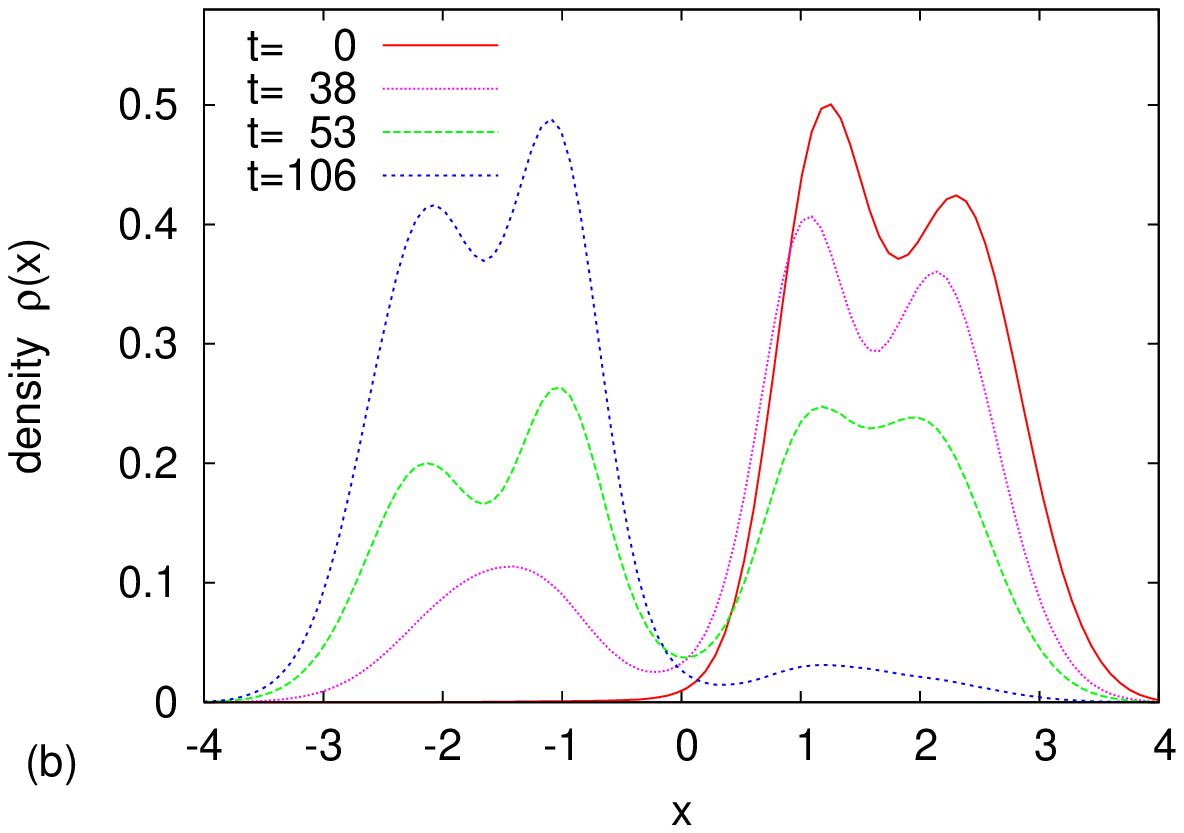}

\caption{(color online) Two-atom dynamics. (a) Population of the right-hand
well over time, $p_{\mathrm{R}}(t)$, for different interaction strengths
$g=0$ (\textcolor{red}{---}), $g=0.2$ (\textcolor{green}{- - -}),
$g=4.7$ (\textcolor{magenta}{${\color{magenta}\cdots}$}), and $g=25$
(${\color{cyan}-\cdot-}$). (b) Snapshots of the one-body density
$\rho(x)$ for different times $t$ in the strongly correlated case
$g=25$. (\emph{All quantities in harmonic-oscillator units throughout,
see text.}) \label{cap:pop-N2}}
\end{figure}

Absent any interactions, the atoms should simply \emph{Rabi-}oscillate
back and forth between both wells. This can be monitored by counting
the percentage of atoms in the right well, $p_{\mathrm{R}}(t)=\langle\Theta(x)\rangle_{\Psi(t)}=\int_{0}^{\infty}\rho(x;t)dx$
($\rho$ being the one-body density) or, correspondingly, the population
imbalance $\delta=p_{\mathrm{R}}-p_{\mathrm{L}}=2p_{\mathrm{R}}-1$.
Figure~\ref{cap:pop-N2}(a) confirms that $p_{\mathrm{R}}$ harmonically
oscillates between 1 and 0. By contrast, if the atoms repel each other,
then the tunneling process will be modified. For $g=0.2$, one sees
that the tunneling oscillations have become a two-mode process: There
is a fast (small-amplitude) oscillation which modulates a much slower
oscillation in which the atoms eventually tunnel completely ($p_{\mathrm{R}}\approx0$).
In case $g$ is increased further, we have found that the tunneling
period becomes indeed so long that complete tunneling may be hard
to observe. For instance, at $g=1.3$ (not displayed here) the period
is as large as $2\times10^{3}$. What remains is a very fast oscillation
with only a minute amplitude -- this may be understood as the few-body
analogue of quantum self-trapping, as will be discussed in Sec.~\ref{sub:Spectrum}.
As we go over to much stronger couplings (see $g=4.7$), we find that
the time evolution becomes more complex, even though this is barely
captured in the reduced quantity $p_{\mathrm{R}}$ {[}Fig.~\ref{cap:pop-N2}(a)].
What is striking, though, is that near the fermionization limit (see
$g=25$) again a simple picture emerges: The tunneling, whose period
roughly equals that of the Rabi oscillations, is superimposed by a
faster, large-amplitude motion. This intriguing result states that
the strongly repulsive atoms coherently tunnel back and forth almost
like a single particle. As an illustration, snapshots of the density
at different $t$ are displayed in Fig.~\ref{cap:pop-N2}(b): At
$t=0$, the \emph{fragmented} pair starts out in the right well, and
gradually tunnels to the left well until the fermionized pair state
reemerges on the left at $t\approx106$.

\subsection{Spectral analysis \label{sub:Spectrum}}

\begin{figure}
\subfigure[]{\includegraphics[width=0.7\columnwidth,keepaspectratio]{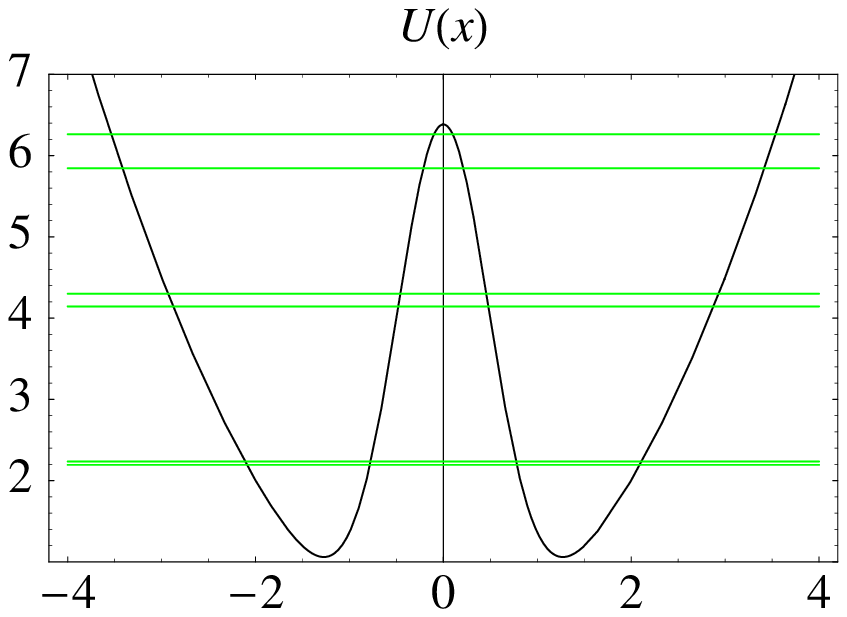}}

\subfigure[]{\includegraphics[width=0.85\columnwidth,keepaspectratio]{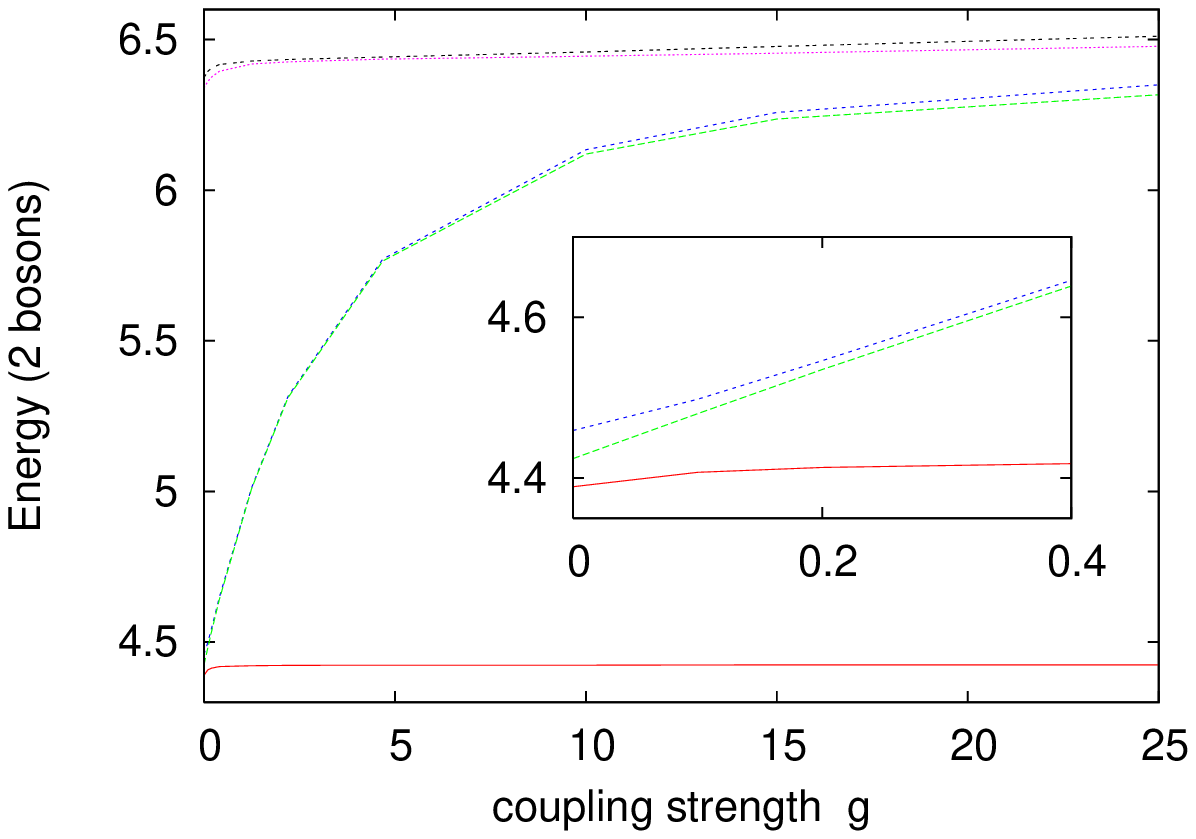}}

\caption{(color online) (a) Single-particle spectrum $\{\epsilon_{a}\}$ of
a double well with barrier height $h=8$. (b) Two-particle spectrum
as a function of the interaction strength $g$. \emph{Inset}: Doublet
formation with increasing $g$. \label{cap:spec-h8}}
\end{figure}

In order to understand the oscillations, let us regard the evolution
of the few-body spectrum $\{ E_{m}(g)\}$ as $g$ is varied (Fig.~\ref{cap:spec-h8}b).
In the noninteracting case, the low-lying spectrum of $N=2$ atoms
is given by distributing all atoms over the symmetric and antisymmetric
single-particle orbital of the lowest doublet (illustrated in Fig.~\ref{cap:spec-h8}a).
This yields the $N+1$ energies \[
\{ E_{m}=E_{0}+m\Delta^{(0)}\mid m=0,\dots,N\},\]
where $\Delta^{(0)}=\epsilon_{1}-\epsilon_{0}$ is the energy gap
between these two orbitals or, in other words, the width of the lowest
\emph{band}. Assuming that for sufficiently small $g$ still only
$N+1=3$ levels are populated in $\Psi(t)=\sum_{m}e^{-iE_{m}t}c_{m}\Psi_{m}$,
then the imbalance $\delta(t)\equiv\langle\Theta(x)-\Theta(-x)\rangle_{\Psi(t)}$
(and likewise $p_{\mathrm{R}}$) can easily be computed to be \begin{equation}
\delta(t)=\delta^{\left(01\right)}\cos(\omega_{01}t)+\delta^{\left(12\right)}\cos(\omega_{12}t),\label{eq:pop-imb}\end{equation}
where $\omega_{mn}=E_{m}-E_{n}$ and $\delta^{(mn)}=4\langle\Psi_{m}|\Theta(x)|\Psi_{n}\rangle c_{m}c_{n}$
is determined by the participating many-body eigenstates. Note that
the term $(mn)=(02)$ vanishes since, by antisymmetry, only opposite-parity
states are coupled. At $g=0$, due to the levels' equidistance, only
a single mode with Rabi frequency $\omega_{01}=\omega_{12}=\Delta^{(0)}$
contributes. For \emph{very small} interaction energies compared to
$\Delta^{(0)}$, the equidistance is slightly lifted, so that the
Rabi oscillations are modulated by a tiny beat frequency $\omega_{01}-\omega_{12}$
(not shown). However, as the interaction is increased further, the
two upper lines $E_{1,2}$ virtually glue to one another to form a
doublet, whereas the gap to $E_{0}$ increases (Fig.~\ref{cap:spec-h8}b,
inset). 

This level adhesion, already calculated for $N\le5$ in Ref.~\cite{zoellner07a},
may be understood from a naive lowest-band two-mode model (see \cite{milburn97}
for details): As $g$ is increased, the on-site interaction energy
eventually overwhelms the tunneling energy $\Delta^{(0)}$, and the
eigenstates evolve from number states $|N_{0}^{(0)},N_{1}^{(0)}\rangle$
in the \emph{delocalized} (anti-)symmetric orbitals $\phi_{a=0,1}^{(0)}$
into superpositions of number states $|N_{\mathrm{L}},N_{\mathrm{R}}\rangle$
in the left/right-\emph{localized} orbitals $\varphi_{\mathrm{L(R)}}^{(0)}=\frac{1}{\sqrt{2}}\left(\phi_{0}^{(0)}\mp\phi_{1}^{(0)}\right)$.
It goes without saying that any two such degenerate number states
$|\nu,N-\nu\rangle\neq|N-\nu,\nu\rangle$ violate parity symmetry
and only serve to form a two-dimensional energy subspace, which for
nonzero $\Delta^{(0)}$ corresponds to the doublets in Fig.~\ref{cap:spec-h8}(b). 

With these considerations on the weak-interaction behavior in mind,
Eq.~(\ref{eq:pop-imb}) asserts that for times $t\ll T_{12}\equiv2\pi/\omega_{12}$,
we only see an oscillation with period $T_{01}\ll T_{12}$, offset
by $\delta^{(12)}$, which on a longer timescale modulates the \emph{slow}
tunneling of period $T_{12}$. For small initial imbalances, we have
$\left|\delta^{(01)}/\delta^{(12)}\right|\propto\left|c_{0}/c_{2}\right|\gg1$;
so for short times we observe the few-body analog of Josephson tunneling.
In our case of an almost complete imbalance, in turn, $|\delta^{(12)}|$
dominates, which ultimately should correspond to \emph{self-trapping},
viz., extremely long tunneling times. These considerations convey
a simple yet \emph{ab initio} picture for the few-body counterpart
of the crossover from Rabi oscillations to self-trapping.

It is obvious that the two-frequency description above breaks down
as the gap to higher-lying states melts (see Fig.~\ref{cap:spec-h8}b),
even though for two atoms no actual crossings with higher states occur,
as opposed to $N\ge3$ \cite{zoellner07a,dounasfrazer07b}. The consequences
for the spectrum are twofold: (i) the quasi-degenerate doublet will
break up again, and (ii) states emerging from higher bands will be
admixed. For the imbalance dynamics, (i) implies that the {}``self-trapping''
scenario will give way to much shorter tunnel periods again, while
(ii) signifies a richer multi-band dynamics. An indication of this
may be seen in Fig.\textbf{~}\ref{cap:pop-N2} for $g=4.7$, but
it most clearly manifests toward fermionization, $g=25$.

In the fermionization limit $g\to\infty$, the system also becomes
integrable again via mapping (\ref{eq:BF}). As an idealization, assume
that at $t=0$ we put two (noninteracting) fermions in the \emph{right-hand}
well, where they would occupy the lowest two orbitals, namely $\varphi_{\mathrm{R}}^{(\beta)}$,
$\beta=0,1$. Expressing this (fermionic) number state $\Psi(0)=\left(\prod_{\beta=0,1}\hat{a}_{\mathrm{R}}^{(\beta)}\right)^{\dagger}|0\rangle$
through the single-particle eigenstates $|\boldsymbol{n}=\{ n_{a_{\beta}}^{(\beta)}\}\rangle_{\!-}$
via $\hat{a}_{\mathrm{R}}^{(\beta)}=\frac{1}{\sqrt{2}}(\hat{a}_{0}^{(\beta)}+\hat{a}_{1}^{(\beta)})$
leads to \[
\Psi(t=0)=\frac{1}{2}\sum_{a_{0},a_{1}\in\{0,1\}}|1_{a_{0}}^{(0)};1_{a_{1}}^{(1)}\rangle_{\!-},\]
where $1_{a_{\beta}}^{(\beta)}$ denotes occupation of the symmetric
($a_{\beta}=0$) or antisymmetric ($a_{\beta}=1$) orbital in band
$\beta$. The frequencies $\omega_{\boldsymbol{n},\boldsymbol{n}'}=E_{\boldsymbol{n}}-E_{\boldsymbol{n}'}$
contributing to $\Psi(t)$ follow in a straightforward fashion:\begin{equation}
\omega_{\boldsymbol{n},\boldsymbol{n}'}=\sum_{\beta,a_{\beta}}\epsilon_{a_{\beta}}^{(\beta)}\left(n_{a_{\beta}}^{(\beta)}-n_{a_{\beta}}^{\prime(\beta)}\right)=\sum_{\beta}\Delta^{(\beta)}\underbrace{\left(n_{1}^{(\beta)}-n_{1}^{\prime(\beta)}\right)}_{=0,\pm1}.\label{eq:fermi-frequ}\end{equation}
 Moreover, let us focus on the imbalance dynamics. Since $\delta^{(\boldsymbol{nn}')}\neq0$
only for opposite-parity states $\boldsymbol{n},\boldsymbol{n}'$,
the sum must contain only an \emph{odd} number of terms. For the special
case of two atoms, we obtain the simple result that the only participating
frequencies are $\Delta^{(0)}$ (the lowest-band Rabi frequency, corresponding
to the longer tunneling period) and $\Delta^{(1)}$ (the larger tunnel
splitting of the first excited band). This links the strongly interacting
dynamics to the noninteracting Rabi oscillations.

\subsection{Role of correlations \label{sub:correlations}}

\begin{figure}
\includegraphics[width=0.85\columnwidth,keepaspectratio]{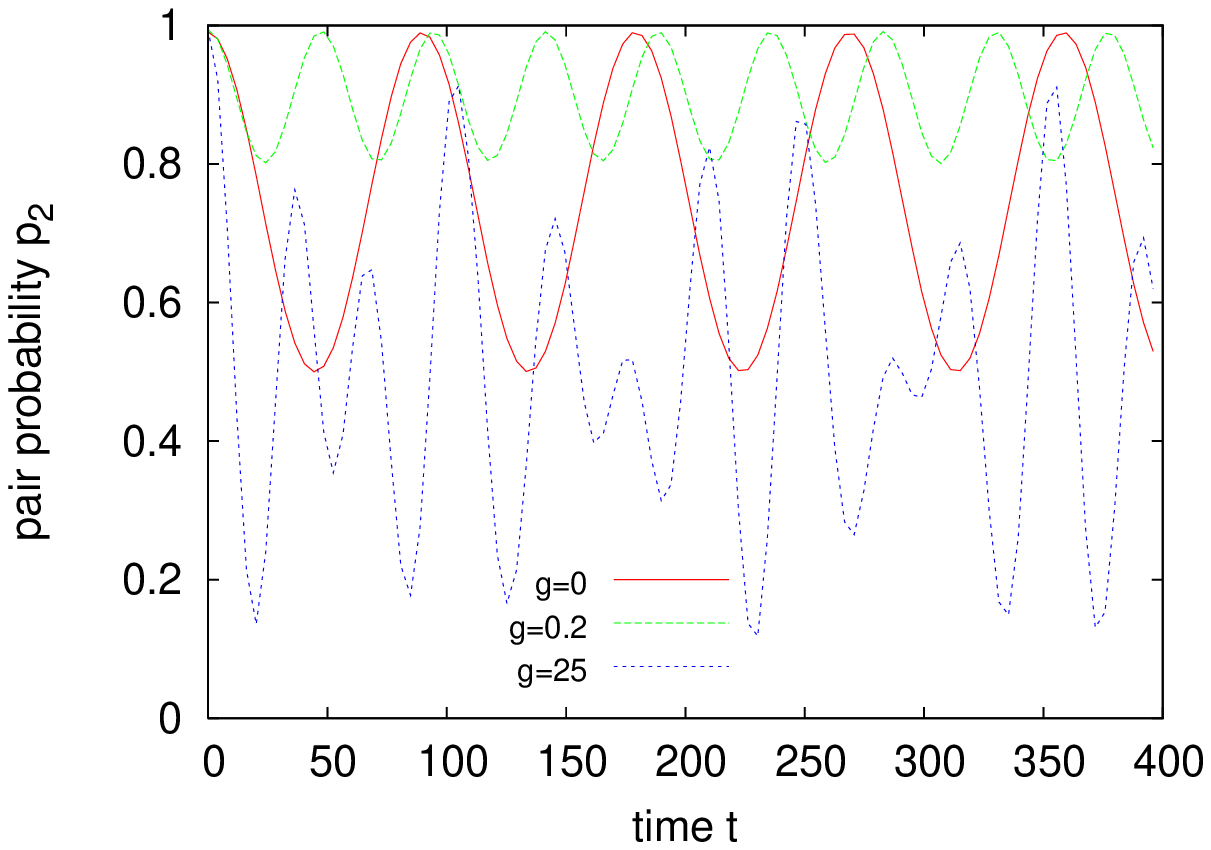}

\includegraphics[width=0.33\columnwidth,keepaspectratio]{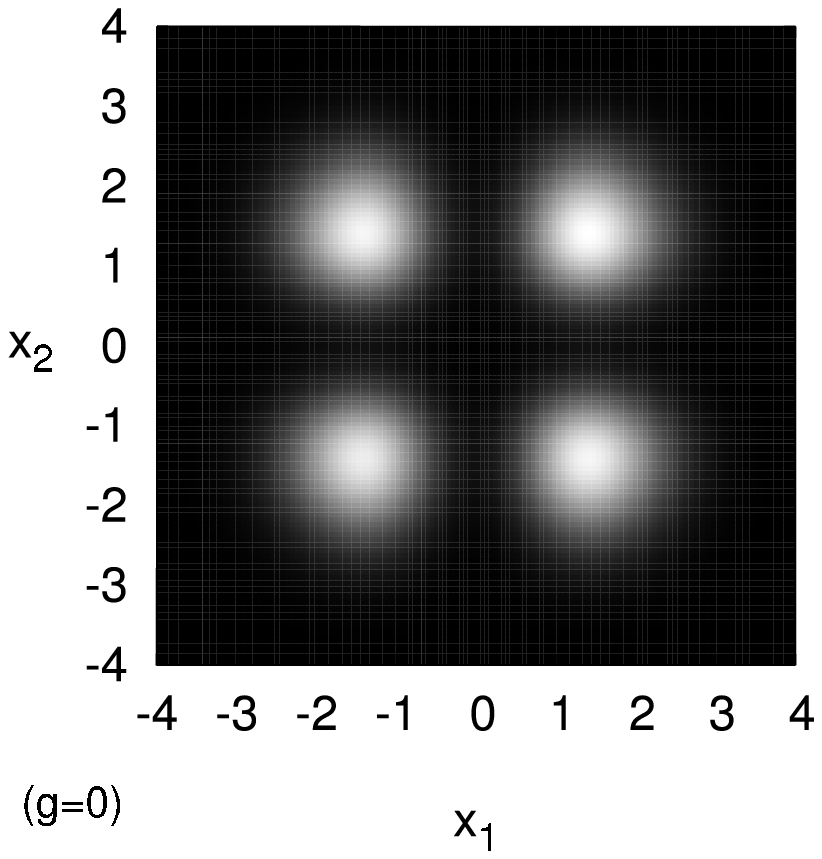}\includegraphics[width=0.33\columnwidth,keepaspectratio]{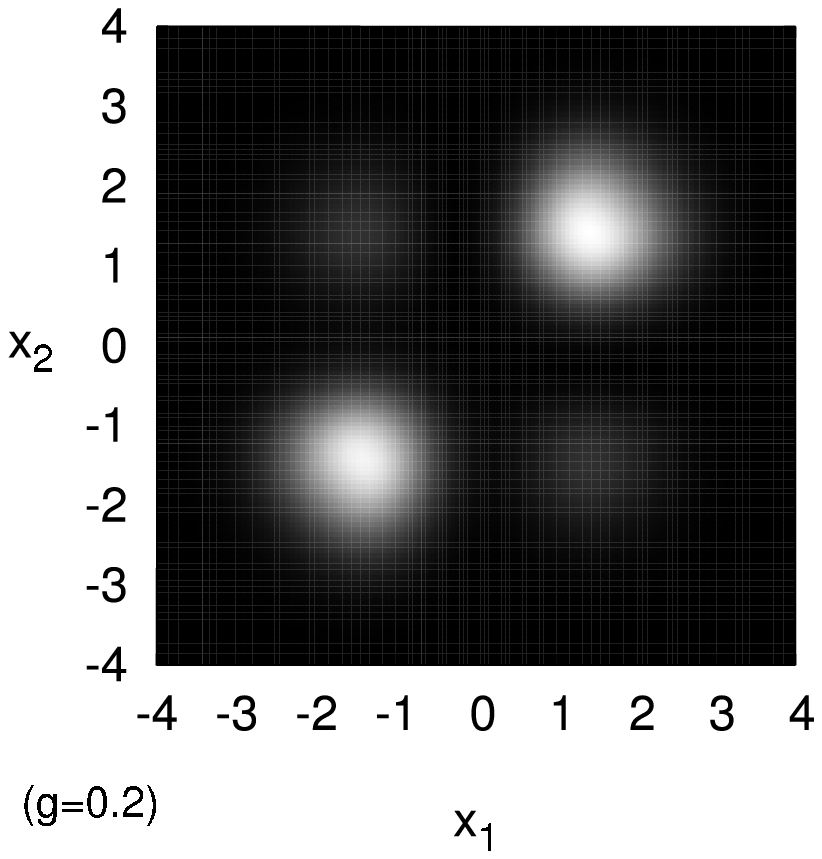}\includegraphics[width=0.33\columnwidth,keepaspectratio]{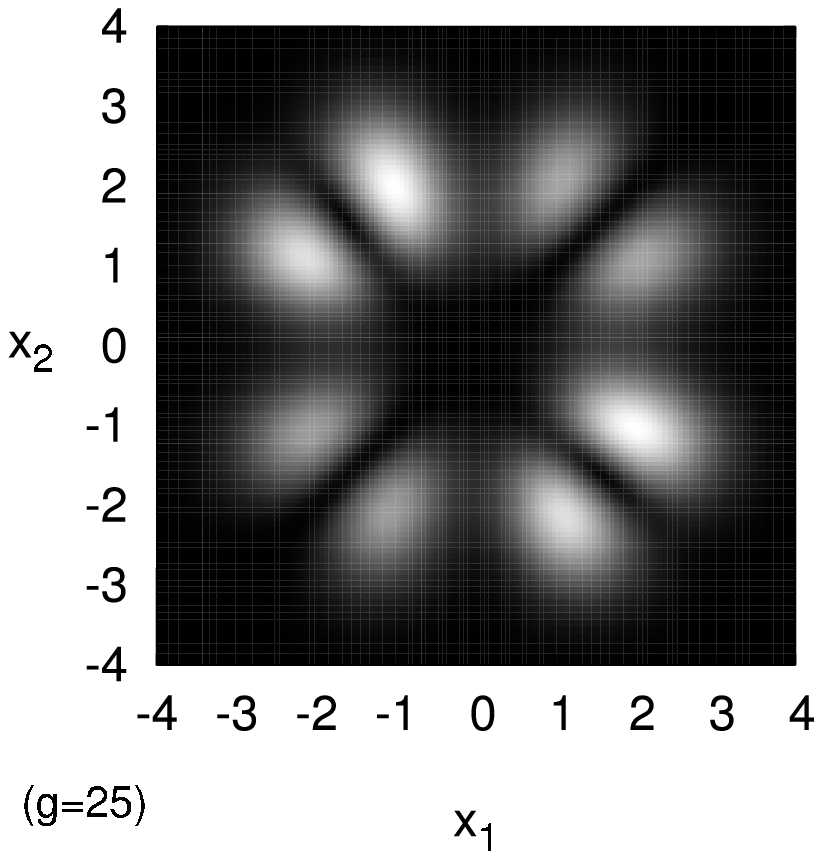}

\caption{(color online) \emph{Top}: Probability $p_{2}(t)$ of finding two
atoms in the same well for $g=0,0.2,25$. \emph{Bottom}: Snapshots
of two-body correlation function $\rho_{2}(x_{1},x_{2})$ at equilibrium
points, $\delta(t_{*})=0$, for $g=0$ ($t_{*}=44$), $g=0.2$ ($t_{*}=128$),
and $g=25$ ($t_{*}=53$) -- \emph{from left to right}. \label{cap:pair}}
\end{figure}

In order to unveil the physical content behind the tunneling dynamics,
let us now investigate the two-body correlations. Noninteracting bosons
simply tunnel independently, which is reflected in the two-body density
$\rho_{2}(x_{1},x_{2})$. As a consequence, if both atoms start out
in one well, then in the equilibrium point of the oscillation it will
be as likely to find both atoms in the same well as in opposite ones.
This is illustrated in Fig.~\ref{cap:pair}, which exposes snapshots
$\rho_{2}(x_{1},x_{2};t_{*})$ at the equilibrium points (where $\delta(t_{*})\stackrel{!}{=}0$)
and visualizes the temporal evolution of the \emph{pair} (or \emph{same-site})
\emph{probability}\begin{eqnarray*}
p_{2}(t) & = & \langle\Theta(x_{1})\Theta(x_{2})+\Theta(-x_{1})\Theta(-x_{2})\rangle_{t}\\
 & = & \int_{\{ x_{1}\cdot x_{2}\ge0\}}\negthickspace\rho_{2}(x_{1},x_{2};t)dx_{1}dx_{2}.\end{eqnarray*}
As we introduce small correlations, the pair probability does not
drop to $0.5$ anymore -- at $g=0.2$ it notably oscillates about
a value near 100\%. This signifies that both atoms can essentially
be found in the \emph{same} well in the course of tunneling, which
is apparent from the equilibrium-point image of $\rho_{2}$. In plain
words, \emph{they tunnel as pairs}. At this point, it is instructive
to revisit the eigenstate analysis above: While the $g=0$ eigenstates
$\Psi_{1,2}$ are \emph{delocalized}, at intermediate $g=0.2$ they
have basically evolved into superpositions $|N_{\mathrm{L}}=2,N_{\mathrm{R}}=0\rangle\pm|0,2\rangle$
of pair states \emph{localized in each well}. In this light, the dynamics
solely consists in shuffling the population back and forth between
these two pair states.

Figure~\ref{cap:pair} in hindsight also casts a light on the fast
(small-amplitude) modulations of $p_{\mathrm{R}}$ encountered in
Fig.~\ref{cap:pop-N2}(a), namely by linking them to temporary reductions
of the pair number $p_{2}$. Thus it is fair to interpret them as
attempted one-body tunneling. Along the lines of the spectral analysis
above, this relates to the contribution from the ground state, in
which the two atoms reside in opposite wells and which does \emph{not}
join a doublet. Since $\Theta(x_{1})\Theta(x_{2})+\Theta(-x_{1})\Theta(-x_{2})$
is parity symmetric, only equal-parity matrix elements contribute
to $p_{2}$, which yields  $p_{2}(t)\approx1-2p^{(02)}\sin^{2}\left(\omega_{02}t/2\right)$. 

It is clear that, as before, the time evolution becomes more involved
as the interaction energy is raised to the fermionization limit (cf.
$g=25$). The two-body correlation pattern is fully fragmented not
only when the pair is captured in one well (corresponding, e.g., to
the upper right corner $x_{1},x_{2}\ge0$), but also when passing
through the equilibrium point $t=53$. These contributions from higher-band
excited states also reflect in the evolution of $p_{2}(t)$, which
is determined by the two modes $\omega_{\pm}=\Delta^{(0)}\pm\Delta^{(1)}$.
Over time, $p_{2}$ passes through just about any value from $1$
(pair) to almost zero (complete isolation). In analogy to free fermions,
it is again tempting to understand this involved pattern as two fermions
tunneling independently with different frequencies.

\subsection{Many-body effects \label{sub:Many-body}}

Although having focused on the case of $N=2$ atoms so far, the question
of higher atom numbers is interesting from two perspectives. For one
thing, at stronger interactions many results become explicitly $N$-dependent,
including distinctions between even/odd atom numbers \cite{zoellner06a,zoellner07a}.
On the other hand, in a setup consisting of a whole \emph{array} of
1D traps like in \cite{paredes04,kinoshita04,kinoshita06}, number
fluctuations may automatically admix states with $N>2$.

\subsubsection{Complete imbalance}

\begin{figure*}
\includegraphics[width=0.49\textwidth,keepaspectratio]{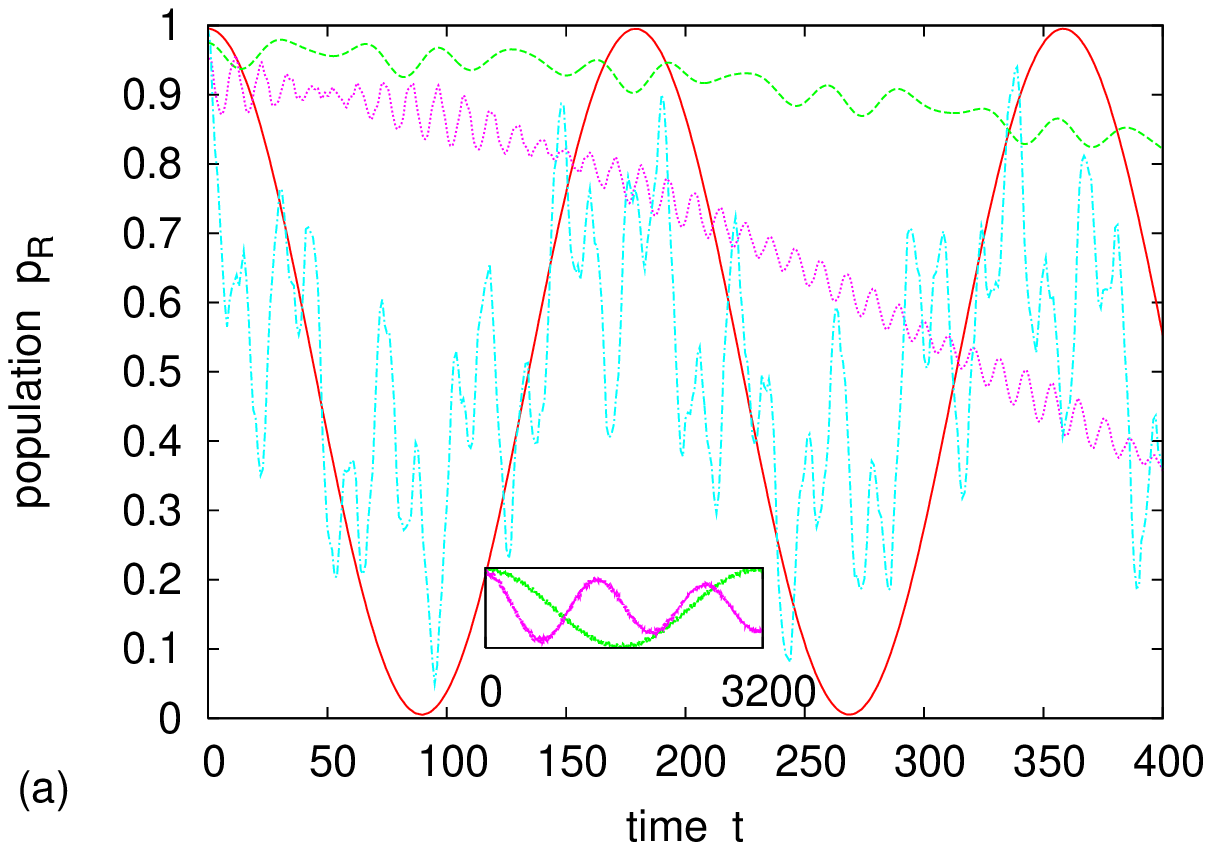}\includegraphics[width=0.49\textwidth,keepaspectratio]{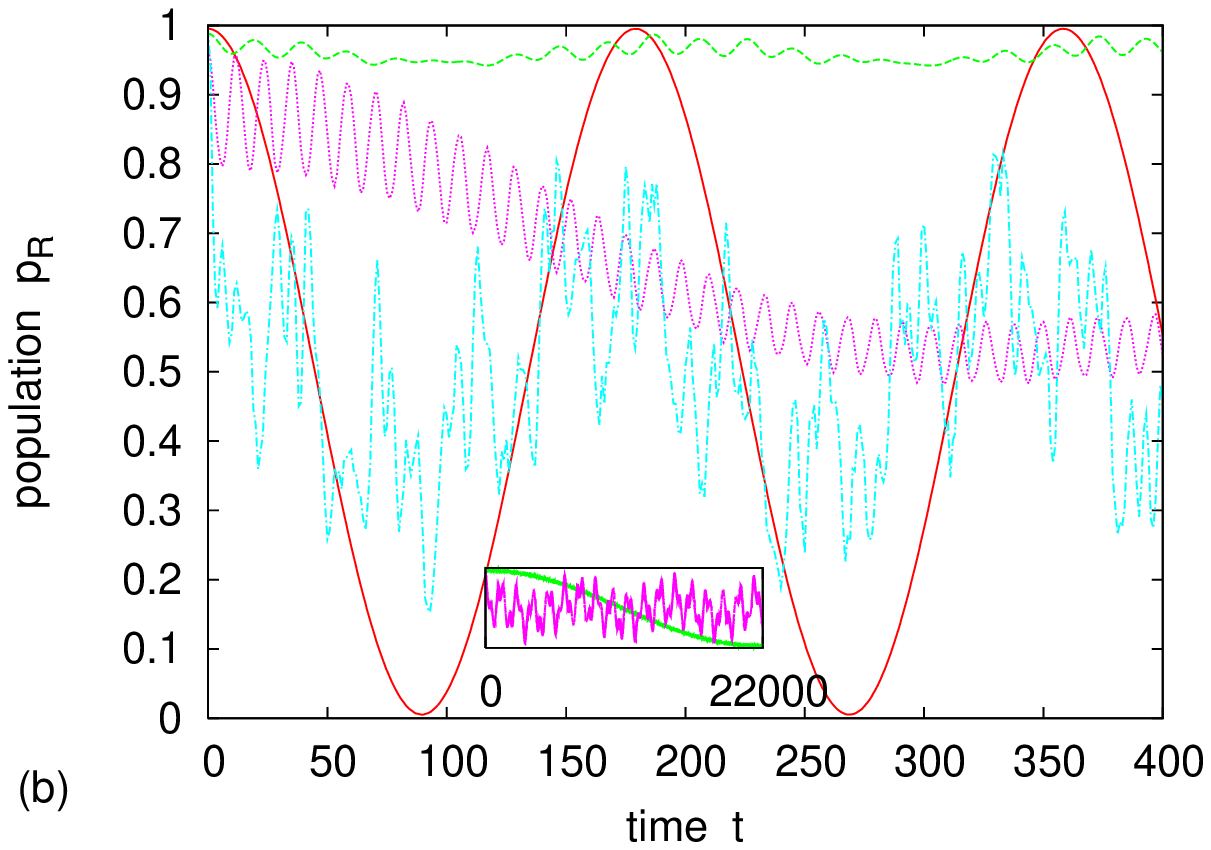}

\caption{(color online) Time evolution $p_{\mathrm{R}}(t)$ of (a) $N=3$,
(b) $N=4$ atoms initially in one well. Shown are the coupling strengths
$g=0$ (\textcolor{red}{---}), $g=0.2$ (\textcolor{green}{- - -}),
$g=4.7$ (\textcolor{magenta}{${\color{magenta}\cdots}$}), and $g=25$
(${\color{cyan}\boldsymbol{-\cdot-}}$). \emph{Insets}: Long-time
behavior for $g=0.2$ (the longer period) and $g=4.7$. (Observe the
different time scales in both insets.) }

\label{cap:N3-4}
\end{figure*}

For $N\ge3$, the weak-interaction behavior does not differ conceptually.
In fact, Eq. (\ref{eq:pop-imb}) carries over, \[
\delta(t)=\sum_{m<n}\delta^{(mn)}\cos(\omega_{mn}t),\]
but with the sum now running over $0\le m<n\le N$. Strictly speaking,
the dynamics is thus no longer determined by two but rather in principle
$N(N+1)/2$ modes $(mn)$ -- although about half of these fail to
contribute by symmetry. Nonetheless, the basic pattern can be understood
from the two-atom case, as will become clear in a moment. 

For $g=0$, assume an ideal initial state with all atoms in the right-localized
orbital $\phi_{\mathrm{R}}=\frac{1}{\sqrt{2}}\left(\phi_{0}+\phi_{1}\right)$
of the lowest band. The weight coefficients $c_{N}(N_{0})=\langle N_{0},N-N_{0}|\Psi(0)\rangle$
with respect to the eigenstates $|N_{0},N_{1}\rangle$ have a binomial
distribution\[
\left|c_{N}(N_{0})\right|^{2}=\frac{1}{2!^{N}}\binom{N}{N_{0}}\stackrel{N\to\infty}{\sim}\delta_{\Delta N_{0}}(N_{0}-\bar{N}_{0})\]
which for larger $N$ asymptotically equals a Gaussian, with a sharp
peak ($\Delta N_{0}=\sqrt{N}/2$) near $\bar{N_{0}}=N/2$. In this
light, only these few states should contribute. Again, the equidistance
of the levels guarantees a simple imbalance oscillation with $\Delta^{(0)}$.
For interaction energies small compared to $\Delta^{(0)}$, the Rabi
oscillations will again be modulated by beats, similar to the case
$N=2$. 

As we move to larger values $g\sim0.2$, the higher-lying of the $N+1$
levels have again merged into doublets \cite{zoellner07a}. In particular,
the highest eigenstate pair was conjectured to be roughly of the form
$|N_{\mathrm{L}}=N,N_{\mathrm{R}}=0\rangle\pm|0,N\rangle$ (in the
limit $h\to\infty$). The idealized state distribution should be peaked
at just these two vectors, whose energy splitting in the bare two-mode
model has been estimated as $\omega\sim2NU/(N-1)!\times(2\Delta^{(0)}/U)^{N}$
\cite{salgueiro06}, where $U$ denotes the on-site interaction energy.
Thus the tunnel period is expected to grow exponentially as $N\to\infty$,
a trend which may be roughly extrapolated from Fig.~\ref{cap:N3-4}
(insets). Ultimately, this should connect to the condensate dynamics
valid for $N\gg1$ \cite{milburn97,tonel05,salgueiro06,creffield07},
when tunneling becomes inaccessible for all intents and purposes.
Of course, realistically, neighboring states will also be excited,
which makes the time evolution richer. However, the separation of
time scales leads to the characteristic interplay of fast, small-amplitude
oscillations (related to attempted single-particle tunneling) and
a much slower tunnel motion, as observed in Fig.~\ref{cap:N3-4}.

Things become more intricate if we leave the two-mode regime, cf.
$g=4.7$. As has been demonstrated in \cite{zoellner07a}, (anti-)crossings
with higher-lying states (which connect to higher-band states at $g=0$)
occur for $N\ge3$. Given our experience of the two-atom case, one
might again expect a simplified behavior as we approach the fermionization
limit. However, we will argue below that this has to be taken with
a grain of salt because an initial state with $N$ hard-core bosons
in one well is highly excited. In the spirit of the Bose-Fermi map,
an idealized state with $N$ fermions prepared in one well will have
contributions from all excitations $|1_{a_{0}}^{(0)};1_{a_{1}}^{(1)};\dots;1_{a_{N-1}}^{(N-1)}\rangle_{\!-}$
($a_{\beta}=0,1$ $\forall\beta$) in the $N$ lowest bands, which
is proven by induction on $N=2$. In view of (\ref{eq:fermi-frequ}),
many more frequencies are expected to be present: Besides the individual
tunnel splittings $\Delta^{(\beta)}$ for each band, these should
in principle be all four combinations $\Delta^{(0)}\pm\Delta^{(1)}\pm\Delta^{(2)}$
for $N=3$, and $4\times4$ combinations $\{\Delta^{(l)}\pm\Delta^{(m)}\pm\Delta^{(n)}\mid0\le l<m<n\le N\}$
for $N=4$ etc, taking into account parity-selection rules. However,
in the fermionization limit with the idealized initial state above,
things simplify even further. Since $\hat{N}_{\mathrm{R}}\equiv\sum_{\beta}a_{\mathrm{R}}^{(\beta)\dagger}a_{\mathrm{R}}^{(\beta)}$
---the Fock-space representation of $\Theta(x)$ in the context of
Eq.~(\ref{eq:pop-imb})--- is a one-particle operator, an eigenstate
$|\boldsymbol{n}\rangle_{\!-}$ is coupled only to {}``singly excited''
states of the type $|\boldsymbol{n}'\rangle_{\!-}=a_{\mathrm{1}}^{(\beta)\dagger}a_{0}^{(\beta)}|\boldsymbol{n}\rangle_{\!-}$
(for some $\beta$), with an excitation frequency $\omega_{\boldsymbol{n},\boldsymbol{n}'}=\Delta^{(\beta)}$.
This yields an imbalance of \[
\delta(t)=\frac{1}{N}\sum_{\beta=0}^{N-1}\cos\Delta^{(\beta)}t\qquad(g\to\infty).\]
This simple formula should be contrasted with the surprising complexity
of the fermionization dynamics already for atom numbers as small as
$N=3,4$. This is illustrated in Fig.~\ref{cap:N3-4}, where $p_{\mathrm{R}}(t)=[(\delta(t)+1]/2$
is plotted (cf. $g=25$). To be sure, for finite $g$ and using a
realistic loading scheme, a few more modes contribute, thus naturally
rendering the dynamics more irregular. But even the inocuous formula
above can account for the seemingly erratic patterns in Fig.~\ref{cap:N3-4}:
The key to see this is to consider the distribution of frequencies
$\{\Delta^{(\beta)}\}$. In the unrealistic limit that $\Delta^{(\beta)}\approx\Delta^{(0)}$
$\forall\beta$, the imbalance would be a neat Rabi oscillation for
any $N$, $\delta(t)\approx\cos\Delta^{(0)}t$. However, a realistic
barrier likely has a Gaussian-type shape and a finite height; hence
the splittings of higher bands tend to grow monotonically. As a consequence,
only the \emph{lower-band} frequencies $\Delta^{(\beta)}$ will contribute
to the tunneling, whereas the higher-band splittings make for much
faster modulations, which average out on a larger time scale. The
gist is that for $N\gg1$, those few lowest-band modes only have a
weight of $O(1/N)$, so in a realistic scenario one expects quasi-equilibration
around $p_{\mathrm{R}}=1/2$.

\subsubsection{Partial imbalance}

\begin{figure}
\includegraphics[width=0.85\columnwidth,keepaspectratio]{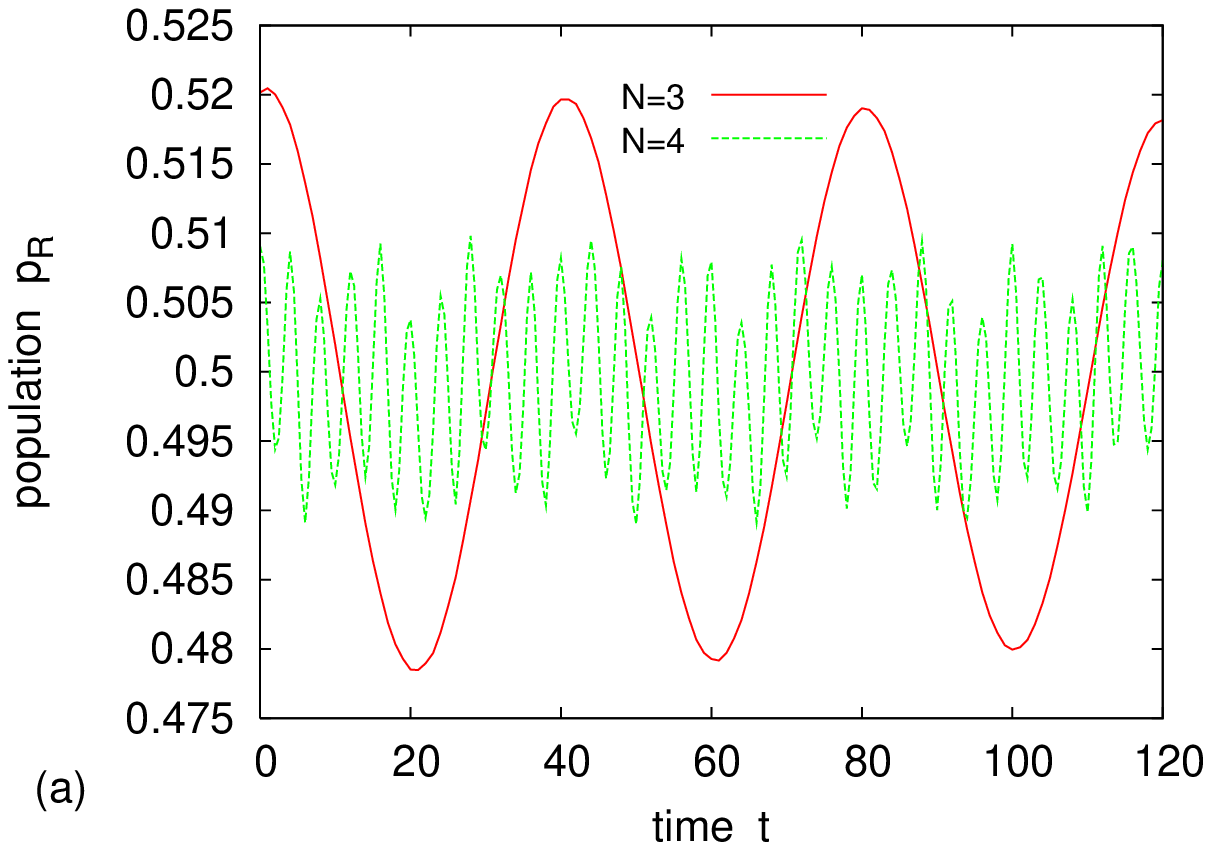}

\includegraphics[width=0.49\columnwidth,keepaspectratio]{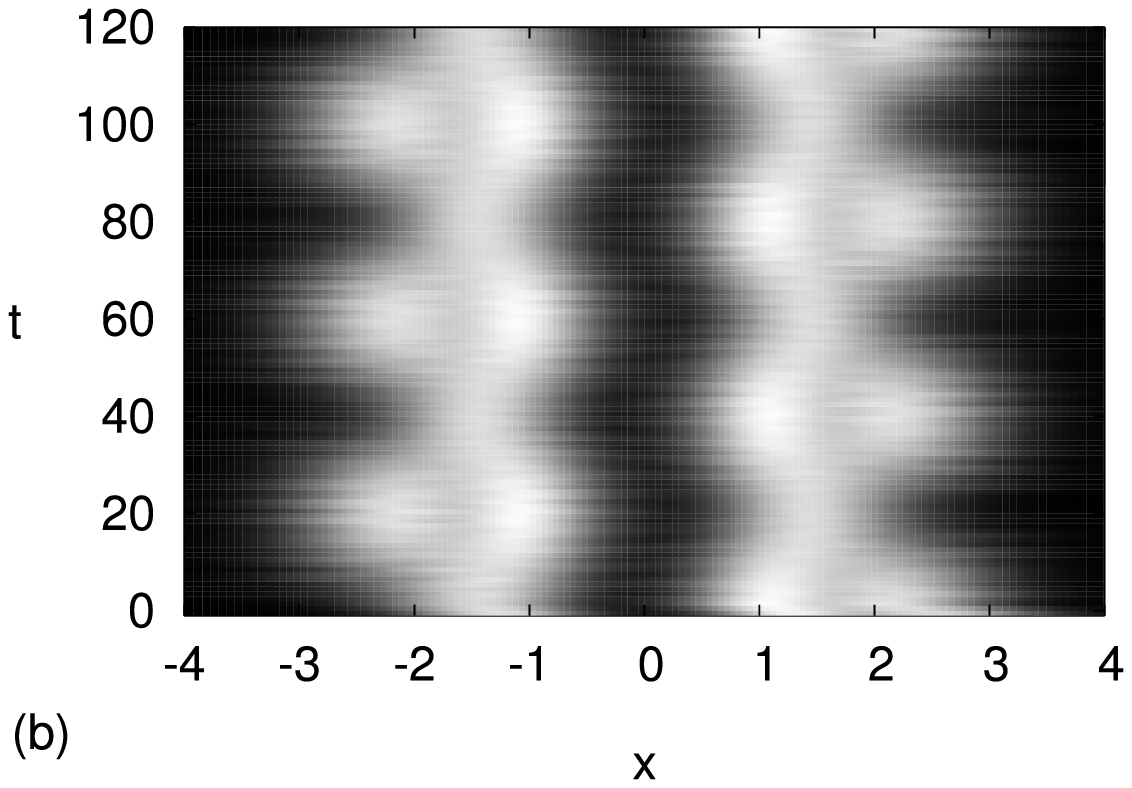}\includegraphics[width=0.49\columnwidth,keepaspectratio]{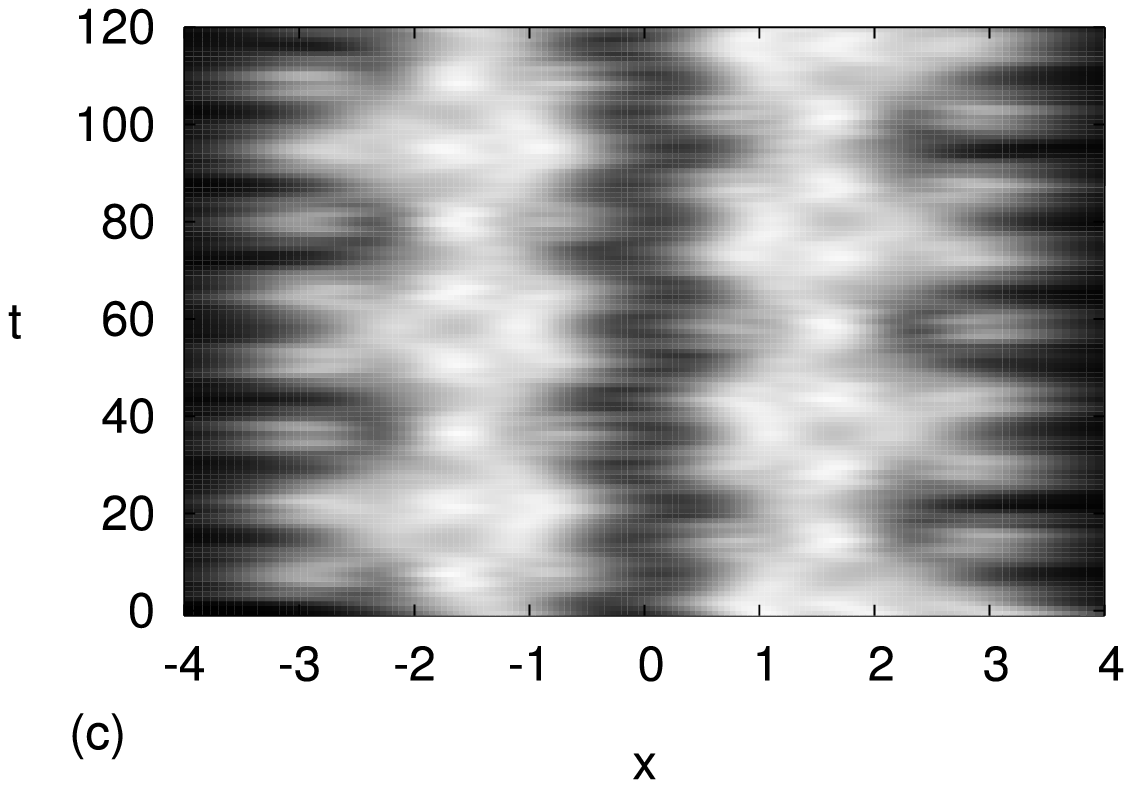}

\caption{(color online) Partial-imbalance effects in the fermionization limit
($g=25$). \emph{}(a) Small-imbalance oscillations (Scenario 1.) for
$N=3,4$ atoms. Plotted is the population of the right-hand well,
$p_{\mathrm{R}}(t)$. \emph{Bottom}: Density evolution $\rho(x;t)$
for $N-1=2$ (b) and $N-1=3$ atoms (c) initially in the right-hand
well if exactly one atom is present on the left (Scenario 2.). \label{cap:N3-4_Jo}}
\end{figure}
While we have so far assumed that all atoms are prepared in one well,
it is natural to ask what the effect of \emph{incomplete imbalances}
$p_{\mathrm{R}}(0)<1$ would be. For simplicity, we will focus on
the fermionization limit (here $g=25$). Two scenarios are conceivable,
in principle:

\begin{enumerate}
\item Small imbalances $p_{\mathrm{R}}\approx1/2$, i.e., small perturbations
of the ground state;
\item Preparing, say, $N-1$ atoms in one well and one in the other.
\end{enumerate}
Option (1.) is plotted in Fig.~\ref{cap:N3-4_Jo}(a) for $N=3,4$.
We clearly observe Josephson-type oscillations in each case, but with
markedly different time scales. This may be understood from the spectral
structure near fermionization \cite{zoellner07a}: For even $N$,
the fermionic ground state $|1_{0}^{(0)},1_{1}^{(0)},\dots,1_{0}^{(N/2-1)},1_{1}^{(N/2-1)}\rangle_{\!-}$
has all bands filled, so that the lowest excitation is created by
moving one atom from band $\beta=N/2-1$ to $\beta=N/2$. Thus the
{}``Josephson'' frequency $\omega_{01}=\epsilon_{0}^{(N/2)}-\epsilon_{1}^{(N/2-1)}$
is a large \emph{inter-band} gap, which for $N=4$ gives a period
of $T_{01}\approx4$. For odd $N$, by contrast, the mechanism is
a different one: Here the ground state leaves the highest band only
\emph{singly} occupied, so that the lowest excitation frequency is
the small \emph{intra-band} splitting $\omega_{01}=\Delta^{(N-1)/2}$.
In Fig.~\ref{cap:N3-4_Jo}(a) ($N=3$), this may be identified as
the rather long period $T_{01}\approx40$.

Scenario (2.), paraphrased in the case $N=3$, is the question of
the fate of an atom pair if the target site (the left well) is already
occupied by an atom. The striking answer, as evidenced in Fig.~\ref{cap:N3-4_Jo}(b),
is that the process can be viewed as single-atom tunneling on the
background of the symmetric two-atom ground state. The tunneling frequency
in the fermionization limit is $\Delta^{(1)}\approx2\pi/40$, which
has the intuitive interpretation of a fermion which---lifted to the
band $\beta=1$---tunnels independently of the two lowest-band fermions.
From that point of view, it should come as no surprise that adding
another particle destroys that simple picture. In fact, Fig.~\ref{cap:N3-4_Jo}(c)
reveals that if we start with $N-1=3$ atoms on the right, then the
tunneling oscillations appear erratic at first glance, and a configuration
with three atoms per site becomes an elusive event (see, e.g., $t\approx22,44$
or $72$). In the fermionic picture, this can be roughly understood
as superimposed tunneling of one atom in the first excited band ($\Delta^{(1)}$)
and another in the second band ($\Delta^{(2)}\approx2\pi/15$), while
the remaining zeroth-band fermions remain inactive.

\section{Asymmetric double well \label{sec:Asymmetric}}

We have so far used the tilt $d$ of the double well merely as a tool
to load the atoms into one well. The question naturally arises whether
the actual tunnel oscillations can be studied in \emph{asymmetric}
wells so as to manipulate the nature of the tunneling. Specifically,
we consider a setup similar to Sec.~\ref{sec:Symmetric}: Two atoms
are prepared in the right well (i.e., in ground state $\Psi_{0}^{(d_{0})}$with
a large initial asymmetry $d_{0}$). Subsequently, the asymmetry is
ramped down to a final value $d\neq0$, thus triggering the tunnel
dynamics.

\subsection{Tuning tunneling resonances}

In symmetric wells, pair tunneling is always resonant in the sense
that an initial state with all atoms on one site is equal in energy
to one with all atoms in the opposite well \cite{dounasfrazer07b,foelling07}.
Conversely, single-atom tunneling should only be likely so long as
the repulsive interaction does not shift the pair state's energy off
resonance with a target state of only a \emph{single} atom on the
left. This squares with our finding that the pair probability $p_{2}$
(Fig.~\ref{cap:pair}) drops to 50\% in the equilibrium points for
$g=0$, while in the correlated case ($g=0.2$) it does not vary considerably
from unity. To condense this insight into a single quantity, let us
define \[
\bar{p}_{1}=\max_{t>0}\{1-p_{2}(t)\}\]
 as the (maximum) \emph{single-atom probability,} relating to the
event of finding the atoms in \emph{different} wells. 

\begin{figure}
\includegraphics[width=0.85\columnwidth,keepaspectratio]{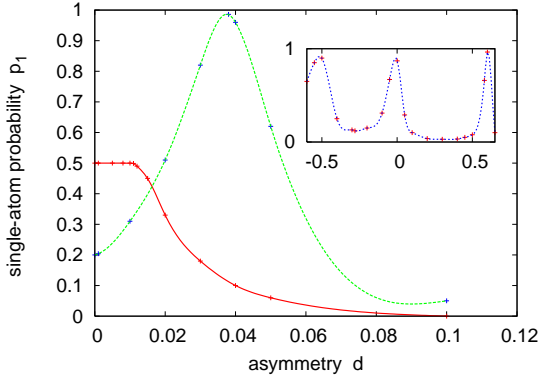}

\caption{(color online) Maximum single-atom probability $\bar{p}_{1}$ as
a function of the tilt parameter $d$. \emph{Solid line:} uncorrelated
tunneling, $g=0$; \emph{dashed line}: correlated tunneling, $g=0.2$.
\emph{Inset}: Near the fermionization limit, $g=25$. Note that the
resonances are not symmetric in $d$ owing to the unsymmetric initial
state $\Psi(0)$. \label{cap:cond-tunn}}
\end{figure}

Figure~\ref{cap:cond-tunn} shows how $\bar{p}_{1}$ changes when
the final asymmetry $d$ between the wells is varied. For $g=0$,
$\bar{p}_{1}(d)$ has a plateau for $d\le0.011$. This relates to
the transition from coexistence of single-atom and pair tunneling
(at $d=0$) to the point where the right-hand well is lowered such
in energy that the initial pair state energetically matches a state
with exactly \emph{one} atom on the left. From the perspective of
the two-body density in Fig.~\ref{cap:pair}, the \emph{final state}
at $d=0.011$ corresponds to the \emph{equilibrium-point} snapshot
for $d=0$. For larger values of $d$, the energy difference between
both wells is too large to transfer a substantial fraction of the
population to the other well.

By contrast, at $g=0.2$ the repulsion is sufficiently strong to drive
the single-atom tunneling off resonance at $d=0$ (Fig.~\ref{cap:cond-tunn}).
Lowering the right well so as to compensate for the interaction-energy
shift leads to a dramatic increase of the tunnel amplitude near $d=0.038$.
The value of $\bar{p}_{1}\approx1$ confirms that this is pure single-atom
tunneling: After half a tunnel period, both atoms are found precisely
in opposite wells, until they return to the pair state on the right
site.

Despite the more convolved dynamics that emerges as we go higher interactions,
the one-atom tunnel resonance persists. However, in the fermionization
limit $g\to\infty$, yet another resonance emerges at $d=0$ already
(Fig.~\ref{cap:cond-tunn}). As in the uncorrelated case, this signifies
coincident single-atom and pair tunneling. This resonance, however,
is much more sensitive to symmetry breaking, which is intelligible
from the picture of two fermions hopping simultaneously in different
bands $\beta=0,1$. Skewing the double well ($d>0$) thus attenuates
both one- and two-atom tunneling until another, pure single-atom resonance
is hit at $d=0.58$. Conversely, energetically \emph{lifting} the
right-hand well ($d\approx-0.5$) makes tunneling to \emph{excited}
target states accessible.

\subsection{Spectral analysis}

To better understand the dependence of the tunnel dynamics on the
tilt $d$, let us consider the two-body spectrum $\{ E_{m}(d)\}$
at fixed coupling $g$. Since both the noninteracting and the fermionization
limit can be deferred from the single-particle spectrum, we will first
stop to review the tilted double well.

\subsubsection{One-body spectrum}

\begin{figure}
\includegraphics[width=0.7\columnwidth,keepaspectratio]{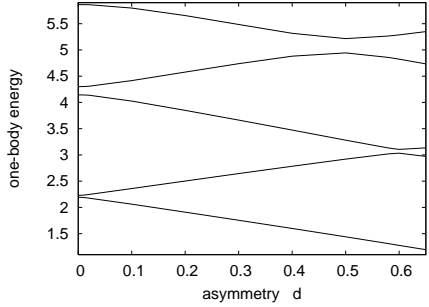}

\caption{Single-particle spectrum $\{\epsilon_{a}(d)\}$ of a double well
as a function of the asymmetry $d$. \label{cap:DWspec_d}}
\end{figure}

Figure \ref{cap:DWspec_d} displays the spectrum $\{\epsilon_{a}(d)\}$
of the double well $U(x)=\frac{1}{2}x^{2}+h\delta_{w}(x)-d\cdot x$
for variable asymmetries $d$. For simplicity, let us resort to a
simple model and expand the one-body Hamiltonian $h(p,x)=\frac{1}{2}p^{2}+U(x)$
in terms of two modes $\phi_{\alpha=\mathrm{L(R)}}$ localized on
the left (right) site (tacitly assuming a fixed band $\beta$). We
denote by 

\begin{itemize}
\item $\langle\phi_{\alpha}|h|\phi_{\alpha}\rangle=\bar{\epsilon}\pm\varsigma/2$
the energies pertaining to \emph{isolated} wells, where the left site
has an energy offset $\varsigma$ 
\item $\left|\langle\phi_{\mathrm{L}}|h|\phi_{\mathrm{R}}\rangle\right|=\Delta/2$
the tunnel coupling.
\end{itemize}
Then a straightforward diagonalization yields \begin{eqnarray*}
\phi_{a,\varsigma} & \propto & \Delta\cdot\phi_{\mathrm{L}}+\left[\varsigma\pm\Delta(\varsigma)\right]\phi_{\mathrm{R}}\qquad(a=0,1)\\
\epsilon_{a,\varsigma} & = & \bar{\epsilon}\mp{\scriptstyle \frac{1}{2}}\Delta(\varsigma)\end{eqnarray*}
where $\Delta(\varsigma)\equiv\sqrt{\Delta^{2}+\varsigma^{2}}$ is
the energy gap in the presence of the tilt. In the symmetric case,
the states are simply given by the (anti-)symmetric orbitals $\phi_{a,\varsigma=0}\propto\phi_{\mathrm{L}}\pm\phi_{\mathrm{R}}$,
with the usual tunnel splitting $\Delta(0)\equiv\Delta$. As we switch
on a tilt $\varsigma>0$, parity is broken and the once delocalized
states break up into one decentered on the left ($\phi_{1}$) and
one on the right ($\phi_{0}$) as $\varsigma\gg\Delta$. This goes
along with a level repulsion of $\epsilon_{0/1,\varsigma}$ about
$\varsigma=0$, where the $\phi_{1}$ state pinpointed on the left
site is energetically lifted, and vice versa. As the states decouple
for $\varsigma\gg\Delta$, the energy approaches that of the isolated
subsystem $\epsilon_{a,\varsigma}\sim\bar{\epsilon}\mp\varsigma/2$.

The above picture holds for each band $\beta$ individually, provided
their levels are well separated. In fact, Fig.~\ref{cap:DWspec_d}
confirms that scenario for tilts small compared to the interband gap,
$\varsigma\ll\bar{\epsilon}^{(\beta+1)}-\bar{\epsilon}^{(\beta)}\sim2$.
For strong enough asymmetries $d$, though, states emerging from different
bands mix, and new avoided crossings are observed in the plot.

\subsubsection{Two-body spectrum}

\begin{figure}
\includegraphics[width=0.7\columnwidth,keepaspectratio]{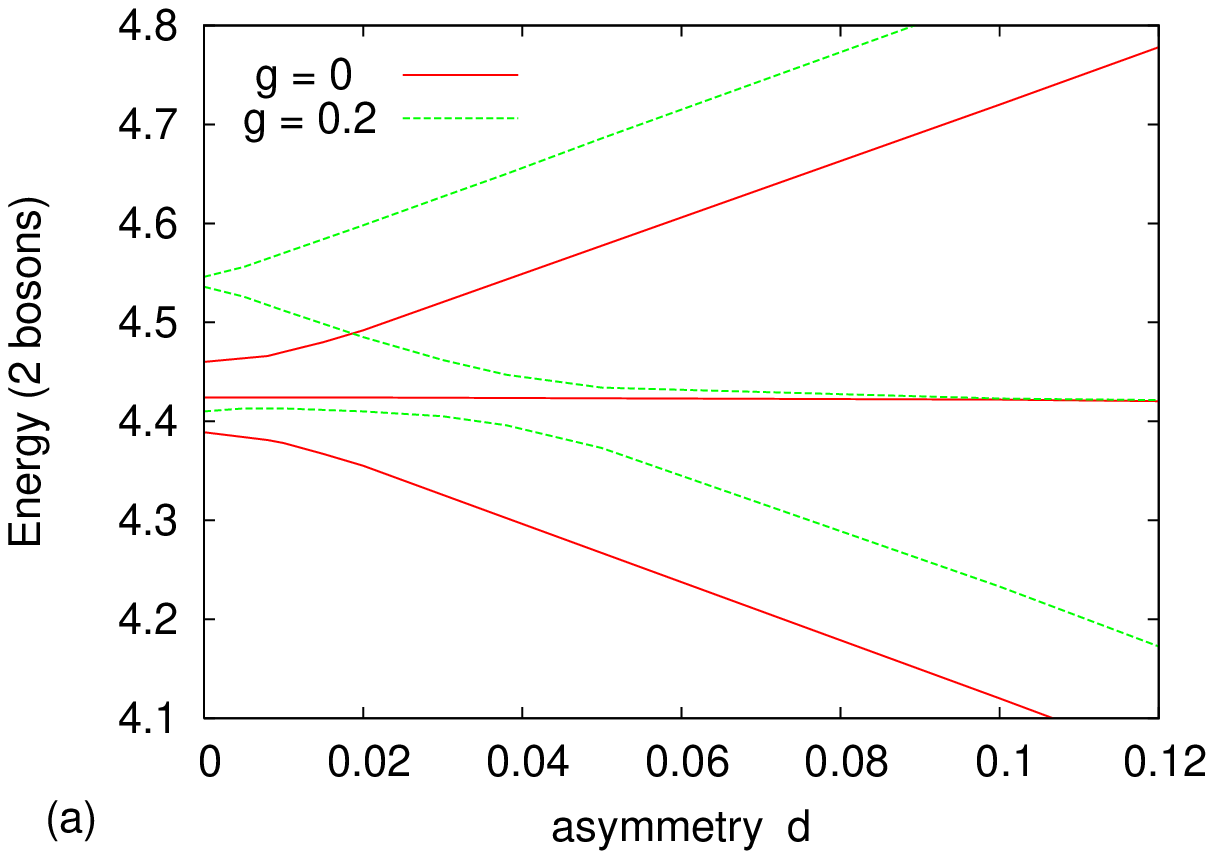}

\includegraphics[width=0.7\columnwidth,keepaspectratio]{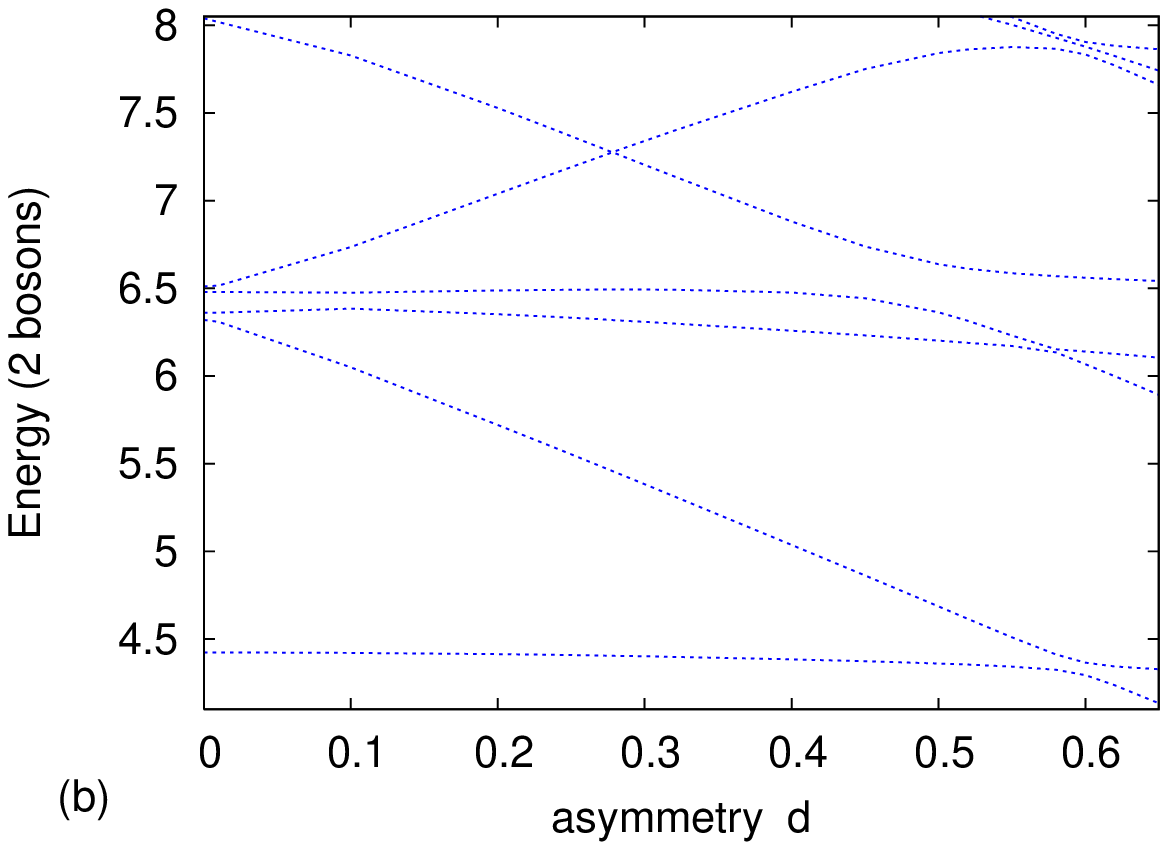}

\caption{(color online) Two-body spectrum $\{ E_{m}(d)\}$ in a tilted double
well. (a) $g=0,\,0.2$ (b) $g=25$. \label{cap:2body-spec_d}}
\end{figure}

\paragraph*{Noninteracting limit:}

In the uncorrelated system, $g=0$, the many-body spectrum $\{ E_{\boldsymbol{n}}=\sum_{a}n_{a}\epsilon_{a}\}$
is obtained from the number states $|\boldsymbol{n}\rangle$ of the
single-particle eigenstates $\phi_{a}$. The energy shift of the levels
$E_{\boldsymbol{n}}(d)$ with respect to $d=0$ thus depends on the
balance between contributions from symmetric orbitals $\phi_{0}^{(\beta)}$
and antisymmetric ones. Specifically, the $d=0$ ground state exhibited
in Fig.~\ref{cap:2body-spec_d}(a) is a coherently symmetric state
$|20\rangle=[\phi_{0}^{(0)}]^{\otimes2}$. Consistently, for perturbations
$d>0$ it localizes on the right, with its level shifting downward
-- in stark contrast to the second excitation $|02\rangle=[\phi_{1}^{(0)}]^{\otimes2}$.
In between, $|11\rangle$ is a compromise between these two borderline
cases in that both partial energy shifts cancel out, leaving a delocalized
state. This gives us a new perspective on the tunneling dynamics reflected
in Fig.~\ref{cap:cond-tunn}. Imagine we start with all atoms prepared
in the right well, viz., the ground state $\Psi_{0}^{(d\to\infty)}$,
and then ramp down $d(t)\to0$ so as to trigger the tunneling. If
we follow the ground-state level nonadiabatically, then at $d=0$
it finds three closely packed levels $E_{m}(0)$ it can couple to
-- in the sense that $\left|\langle\Psi_{m}|\Psi(0)\rangle\right|\neq0$,
so that a nontrivial dynamics becomes possible. In fact, at $d=0$,
these correspond to Rabi oscillations. If we were to choose a final
asymmetry $d\stackrel{<}{\sim}0.01$ (in the notation above, $\varsigma^{(0)}<\Delta^{(0)}$),
roughly the same level would be available, confirming the plateau
encountered in Fig.~\ref{cap:cond-tunn}. However, for final values
$d>0.01$, the levels decouple, and no longer are there any target
states at disposal for tunneling.

\paragraph*{Medium interactions:}

These elementary thoughts also help us explore the nontrivial dynamics
for intermediate couplings, as shown for $g=0.2$ in Fig.~\ref{cap:2body-spec_d}(a).
The $d=0$ ground state, in the limit $\Delta^{(0)}\to0$, has the
Mott-insulator form $|1_{\mathrm{L}}1_{\mathrm{R}}\rangle$ and should
be insensitive to symmetry breaking $d>0$. By contrast, the quasi-degenerate
excited pair $|2_{\mathrm{L}}0_{\mathrm{R}}\rangle\pm|0_{\mathrm{L}}2_{\mathrm{R}}\rangle$
only requires a minute perturbation to break up into two localized
states. It is plain to see that, at $d\approx0.04$, the lower excited
curve anti-crosses the ground state, and the two states are virtually
swapped. Resorting again to a simple two-mode model, the (avoided)
crossing occurs for tilts $\varsigma=U$ matching the on-site repulsion
energy. 

The bearing this has on the tunnel dynamics is evident: Apart from
the self-trapping scenario at $d=0$, there is a fairly broad tunnel
resonance at $d\approx0.04$, where the fully imbalanced initial state
$\Psi(0)$ couples to that with one atom on each site, $|1_{\mathrm{L}}1_{\mathrm{R}}\rangle$.
This is but the one-body resonance encountered in Fig.~\ref{cap:cond-tunn}.
To come by a crude estimate for the critical value $d_{c}$, assume
that the energy of initial and final states match, $\langle H_{d_{c}}\rangle_{\mathrm{i}}=\langle H_{d_{c}}\rangle_{\mathrm{f}}$.
Modeling the initial pair state by the ground state $\Psi_{0}^{(d_{0})}$
(at the initial $d_{0}>0$), and the final state with a single atom
on the left by $\Psi_{0}^{(0)}$, yields the estimate\[
d_{c}=d_{0}-\left(E_{0}^{(0)}-E_{0}^{(d_{0})}\right)/N\left\langle x\right\rangle ^{(d_{0})}\]
in terms of the ground-state energies at the initial $d_{0}>0$ and
$d=0$, respectively, and the elongation $\left\langle x\right\rangle $
at time $t=0$.

\paragraph*{Fermionization limit:}

Figure~\ref{cap:2body-spec_d}(b) shows the spectrum near fermionization,
$g=25$. The $d=0$ ground state turns out to be widely robust against
perturbations, which can be understood from the fact that its fermionic
counterpart $|1_{\mathrm{0}}^{(0)}1_{1}^{(0)}\rangle_{\!-}$ has balanced
populations of right- and left-localizing orbitals. The only way to
obtain a \emph{right-localized} ground state is to lower one well
enough for it to hit a localized state from the upper band. This
is what happens at $d\approx0.6$, where the tilt energy $\varsigma^{(1)}/2=\bar{\epsilon}^{(1)}-\bar{\epsilon}^{(0)}$
compensates the inter-band gap. That crossing marks just the one-body
resonance seen in Fig.~\ref{cap:cond-tunn} at $d\approx0.6$. In
the fermionic picture invoked above, it may be thought of as one \emph{excited}
fermion tunneling to the \emph{lowest} level on the left.

If we follow the localized state nonadiabatically, then at $d=0$
we recover the mixed single-atom/pair resonance laid bare in Fig.~\ref{cap:cond-tunn}.
Further ramping up the right well to $d\approx-0.3$ (where the spectrum
is mirrored at $d=0$), we see yet another crossing. A closer look
reveals that the partner state is entirely localized on the \emph{left},
so that one might hope for a pair resonance. However, as both states
are localized in disjoint regions, they are not coupled by the perturbation
($-d\cdot x$), and in practice no tunnel resonance is observed. It
may be illuminating to look at this from the fermionic perspective.
For $d\approx-0.3$, the initial state on the right is $\Psi(0)\approx|1_{1}^{(0)}{;1}_{1}^{(1)}\rangle_{\!-}$,
while the partner state emanating from $E(0)\approx8$ in turn is
given by $|1_{0}^{(0)}{;1}_{0}^{(2)}\rangle_{\!-}$. In this light,
the tunneling {}``resonance'' in question refers to the following
situation: Two fermions simultaneously hop from the zeroth (first
excited) level on the right down to the zeroth level (up into the
second level) of the energetically lower left site. While both processes
individually are off resonance, the total energy is conserved. This
reflects in the one-body spectrum (Fig.~\ref{cap:DWspec_d}), where
no avoided crossing is to be observed at $d\approx-0.3$ -- rather,
there is an accidental crossing of the sums $E_{\boldsymbol{n}}=\sum_{a}n_{a}\epsilon_{a}$.
However, at $d\approx-0.6$, another \emph{avoided} crossing emerges,
which---in the fermion language---corresponds to multiple one-body
resonances with the first and second excited level in the left well.

\section{Conclusions and outlook}

We have analyzed the crossover from uncorrelated to fermionized tunneling
of few 1D bosons in a double well. The pathway leads via strongly
delayed pair tunneling for medium interactions---associated with doublet
formation in the few-body spectrum---to fermionized tunneling, where
the strongly correlated atoms tunnel back and forth with characteristic
modulations. By analogy to free fermions, these may be understood
as multi-band Rabi oscillations, which become more and more complex
and quasi-equilibrate for large atom numbers. To uncover the physical
mechanisms, it is essential to study two-body correlations. These
reveal a strong suppression of single-atom tunneling for intermediate
coupling, with a revival toward fermionization, where an involved
interplay of pair and single-atom tunneling is observed. 

Whereas for small interactions, higher atom numbers essentially only
increase the tunnel period but do not change the scenario qualitatively,
the multi-atom dynamics becomes much richer as fermionization is approached.
Apart from the above case of a \emph{complete} initial imbalance,
this applies to situations where not all atoms are initially in one
well. In particular, Josephson-type small-amplitude oscillations exhibit
vastly different time scales for odd/even numbers. On the other hand,
initially storing an extra atom in the \emph{target} well suppresses
the lowet-band tunneling and thus leads to a simplified dynamics.

Finally, studying the dynamics in \emph{asymmetric} wells provides
a valuable perspective on the tunnel mechanism in terms of one- and
two-atom tunnel resonances. Depending on the energy difference between
the sites, the tunnel amplitude can be largely enhanced or suppressed.
For noninteracting bosons, this has been described by a plateau of
the single-atom probability about the asymmetry parameter $d=0$.
At medium interactions, in turn, single-particle tunneling becomes
resonant only when the energy offset of one well compensates the interaction-energy
shift at $d>0$. In the fermionization limit, another $d=0$ resonance
emerges, accompanied by higher-level resonances at $d\neq0$. Those
features are explained in terms of avoided crossings in the spectrum
as $d$ is varied. Such a deeper understanding of the tunneling may
pave the way to an active control of strongly correlated systems,
for instance by allowing to transport definite numbers of atoms from
a reservoir to a target well.

\begin{acknowledgments}
Financial support from the Landesstiftung Baden-Württemberg in the
framework of the project {}`Mesoscopics and atom optics of small
ensembles of ultracold atoms' is acknowledged by P.S. and S.Z. The
authors also thank L.~D. Carr, S.~Jochim, and C.~H.~Greene for
fruitful discussions.
\end{acknowledgments}
\bibliographystyle{prsty}
\bibliography{mctdh,phd}

\end{document}